\documentclass[12pt]{article}
\usepackage{graphicx}
\usepackage{epsfig}
\usepackage{bm}
\begin{document}
\title{Recent Models of Neutrino Masses and Mixing}
\author{Ambar Ghosal\\
Saha Institute of Nuclear Physics,\\
1/AF Bidhannagar, Kolkata 700 064, India}
\date{\today}
\maketitle
\section{Introduction}
Neutrino physics is now playing a major role to probe physics 
beyond standard model. Confirmation of tiny but non-zero neutrino 
mass through neutrino oscillation experiments have thrown light on the 
structure of leptonic sector. On neutrino mass, there are some direct 
experimental upper bounds on the individual neutrino masses from the 
decay of other particles, nuclei and also on the sum of the masses 
of neutrino from cosmological observations whereas neutrino oscillation 
experiments put bounds on neutrino mass-squared differences on neutrinos 
which oscillate into each other. Furthermore, there is an additional 
bound on the effective mass of the Majorana type neutrino mass coming 
from the $\beta\beta_{0\nu}$ decay. This bound essentially restricts 
$11$ element of the neutrino mass matrix. 
\section{Experimental Bounds}
\subsection{Direct bounds}
A direct bound on $m_{\nu_\tau}$ is obtained from the hadronic decay
of $\tau$ as $m_{\nu_\tau}<18$MeV ($95\%$ C.L.)  and from the $\pi$
decay a bound on $m_{\nu_\mu}$ is coming as $m_{\nu_\mu}<170$KeV
($95\%$ C.L.), and from the tritium beta decay a bound on $m_{\nu_e}$
is coming as $m_{\nu_e}<2.2$eV ($95\%$ C.L.). Future KATRIN experiment
\cite{data1} will bring down this limit to $m_{\nu_e}<0.2$eV.
\subsection{Indirect Bounds}
Bounds on solar neutrino mass splitting if $\nu_e$ is oscillated to
$\nu_\mu$, combining SNO+KamLAND+SAGE+GALLEX \cite{solar,kl2} neutrino
experimental results give $\Delta m^2_{21} = (7.3-8.5)\times{10}^{-5}$
$\rm eV^2$ at $2\sigma$ range with the best fit value $\Delta m^2_{21}
= 7.9\times{10}^{-5}$ $\rm eV^2$ \cite{data2,analyzes} and the bound
on solar neutrino mixing angle $\theta_{12}$ is $\sin^2\theta_{12}$=
$0.3$ with the $2\sigma$ range is 0.26 -
0.36\cite{data2,analyzes,Strumia:2005tc}.  The global analysis of
Atmospheric + K2K \cite{SKatm,k2k} experimental results give the
best-fit value of $\Delta m^2_{31} = 2.5\times{10}^{-3}$ $\rm eV^2$
and the $2\sigma$ range give the upper and lower bounds as (2.1
-3.0)\cite{Strumia:2005tc,data2}. Including MINOS experimental data
the best fit is shifted to $\Delta m^2_{31} = 2.2\times {10}^{-3}$
$\rm eV^2$\cite{data2}. The determination of $\theta_{23}$ mixing
angle is completely dominated by atmospheric data and there is no
change due to MINOS data. The best fit value of $\sin^2\theta_{23}$=
$0.5$ with the $2\sigma$ range 0.38 - 0.64. The CHOOZ
\cite{Apollonio:2002gd} experimental upper bound on $\theta_{13}$ is
given by $\theta_{13}<12^o$ (at 3$\sigma$) with best fit value
$\sin^2\theta_{13}\simeq 0.0$ .  An additional bound on the sum of the
neutrino masses is coming from the observation of WMAP \cite{wmap}
experimental result on cosmic microwave background
anisotropies. Combining analysis of 2dF Galaxy Redshift survey, CBI
and ACBAR \cite{Elgaroy:2002bi}, WMAP has determined the amount of
critical density contributed due to relativistic neutrinos which in
turn gives an upper bound on the total neutrino mass as $\Sigma
m_i\leq 0.7$ eV at $95\%$ C.L . Another bound is due to
$\beta\beta_{0\nu}$ decay which gives the bound on effective Majorana
neutrino mass as $<|m_{ee}|>$ = (0.05 -0.84) eV at $95\%$ C.L
\cite{beta}(relaxing the uncertainty of the
nuclear matrix elements up to $\pm 50\%$ and the contribution to this
process due to particles other than Majorana neutrino is negligible).
\section{Models with $\theta_{23}$ = $45^o$ and $\theta_{13}$ = $0^o$}
At the lowest order, models with maximal atmospheric mixing angle, 
($\theta_{23} = 45^o$) and vanishing value of CHOOZ mixing angle 
($\theta_{13} = 0$) driven by some symmetry in addition with the 
standard model obtained through the choice of following neutrino  
mass matrix assuming the charged lepton mass matrix diagonal. 
Consider the following neutrino mass matrix 
\begin{equation}
m_\nu = \pmatrix{a&b&b\cr
                 b&c&d\cr
                 b&d&c}
\end{equation}
Upon diagonalization, the above mass matrix gives three non-degenerate
eigenvalues and two out of the three mixing angles are independent of
model parameters and they are ($\theta_{23} = 45^o$), ($\theta_{13} =
0$), however, the solar mixing angle depends upon the model
parameters.  \par The above mass matrix can easily be obtained
invoking $\mu$ - $\tau$ reflection symmetry 
\cite{mutau,Ghosal:2004qb} and regarding
that symmetry a point should be noted. If we invoke
$l_{2L}\leftrightarrow l_{3L}$ and $\mu_R\leftrightarrow \tau_R$ then
a diagonal charged lepton mass matrix will lead to unacceptable
relation $m_\mu = m_\tau$. Thus it is necessary to build up a concrete
model keeping track of the contribution from the charged lepton
sector.  \par There is another approach to generate the above mixing
angles through non-diagonal charged lepton mass matrix and keeping
neutrino mass matrix diagonal. Examples of such types are $S_3\times
S_3$ \cite{s3} permutation symmetry which gives rise to 'democratic
type' charged lepton mixing matrix, lopsided structure of charged
lepton mass matrix using GUT model \cite{gut}.

Apart from the discrete symmetries few continuous $U(1)$ symmetries
are important in the context of neutrino mass and mixing pattern.
Among them $U(1)_{L_e-L_\mu-L_\tau}$ \cite{lelmlt} is the most
popular one. The neutrino mass matrix in the basis of diagonal charged
lepton mass matrix this symmetry gives
\begin{equation}
m_\nu =\pmatrix{0&p&q\cr
                p&0&0\cr
                q&0&0}. 
\end{equation}
Along with the $\mu-\tau$ exchange symmetry (implied $p=q$) 
it generates $\theta_{13} = 0$, 
$\theta_{12} = \theta_{23} = 45^o$
This is so called bimaximal mixing pattern which is the consequence 
of those flavor symmetries, however, $\theta_{12} = 45^o$ 
goes beyond the $5\sigma$ value of the present experimental solar 
neutrino data.  
There is another $U(1)$ symmetry which 
can generate $\theta_{13} = 0$, $\theta_{23} = 45^o$, namely, 
$U(1)_{L_\mu-L_\tau}$ symmetry \cite{lmmlt}, which generates quasi-degenerate 
mass spectrum of neutrinos. 
In addition it generates $\theta_{12}=0$ which can be cured under 
soft breaking of the symmetry.
\section{Models with $\theta_{23}\simeq 45^o$ and $\theta_{13}\neq 0^o$}
In this section, we discuss a specific model which results the 
above mentioned mixing angles, however, still arbitrary solar 
mixing angle. The model is based on $\mu-\tau$ reflection symmetry 
as follows:\\
\begin{equation}
l_{2L}\leftrightarrow l_{3L}\quad\quad  {\rm and}
\quad\quad  \mu_R\leftrightarrow \tau_R
\end{equation}     
\noindent
Neutrino masses are generated through higher dimensional operators,
such as , $ll\phi\phi$/M where $\phi$ is the usual Higgs doublet.  The
resulting charged lepton and neutrino mass matrices are
\cite{Ghosal:2004qb}
\begin{equation}
m_E =\pmatrix{a&b&b\cr
              c&d&e\cr
              c&e&d}, 
m_\nu =\pmatrix{p&q&q\cr
                q&r&r^\prime\cr
                q&r^\prime&r}
\end{equation}
Considering both the mass matrices are complex upon diagonalization 
the resulting mixing angles are coming out as 
\begin{equation}
\sin^2\theta_{atm}\simeq 0.99,\,\, 
\sin^22\theta_{13} = O(10^{-5}),\,\,
\sin^2\theta_{solar}\simeq 0.85
\end{equation}
for a specific choice of model parameters with acceptable charged
lepton and neutrino mass values. There are several other types of
models exist in the literature \cite{u13mutau} which results to the
above value of mixing angles.For all those types of models, solar
mixing angles  depends on the choice of model parameters. Thus,
our search for a more predictive model driven by some symmetry is
still in progress.

\section{Tri-bimaximal Mixing}
Experimental values of the three neutrino mixing angles are found to be 
remarkably close to the conjectured tri-bimaximal mixing ansatz proposed 
by Harrison, Perkins and Scott (HPS) \cite{tb} is given by 
\begin{equation}
U_{tb}=\pmatrix{\sqrt{2 \over 3} & \sqrt{ 1 \over 3} & 0 \cr
-\sqrt{ 1 \over 6} & \sqrt{ 1 \over 3} & - \sqrt{ 1 \over 2} \cr
-\sqrt{ 1 \over 6} & \sqrt{ 1 \over 3} & \sqrt{ 1 \over 2}
}.
\end{equation}
which gives $\theta_{12} = 35.2^o$, $\theta_{23} = 45^o$,
$\theta_{13} = 0^o$ with a suitable normalization of the signs for the 
matrix elements. This mixing pattern has been studied by several authors.
Many theoretical efforts have been made to generate such mixing naturally 
dictated by some symmetry. A successful candidate is the Non-Abelian 
discrete $A_4$ symmetry through the incorporation of which with the 
standard model the HPS ansatz is realized. In the next section we will 
present a brief introduction to $A_4$ symmetry. To understand the 
very different mass pattern of charged lepton and neutrino 
$A_4$ symmetry has been proposed by E.Ma and others \cite{ma}. However, 
to realize the HPS mixing within the framework of an 
$SU(2)_L\times U(1)_Y\times A_4$ model has been done by 
Altarelli and Feruglio \cite{Altarelli1}. 
\section{$A_4$ Symmetry}
$A_4$ is a non-abelian discrete symmetry group generated due to even 
permutation of four objects. It can also be viewed as an equivalent 
representation of the symmetry of a Tetrahedron. Number of elements of 
$A_4$ is 12 and those elements can be divided in four distinct classes as 
follows : \\
$\bullet$ Identity transformation : E(1)\\
$\bullet$ Four three fold axis of rotation : $C_3$(4)\\
$\bullet$ Four three fold axis of rotation : $C_3^2$(4)\\
$\bullet$ Three two fold axis of rotation  : $C_2$(3)\\
where $C_n = e^{2\pi i/n}$
\par
Obviously, $A_4$ has four irreducible representation with dimension 
${\Sigma}_{i=1}^4 n_i^2$ = 12 and the only solution exists as 
\begin{equation}
n_1 = n_2 = n_3 =1 , n_4 = 3
\end{equation}
Those irreducible representations are labeled as 
$1$, $1^{'}$, $1^{''}$ and $3$. Multiplication between different 
representations are given by 
\begin{flushleft}
3$\times$ 3=1+$1^{'}$+$1^{''}$+3+3\\
1$\times$1=1, \quad 1$\times$$1^{'}=1^{'}$, \quad 1$\times$
$1^{''}$=$1^{''}$\quad $1^{'}$$\times$$1^{'}$=$1^{''}$, 
\quad $1^{'}$ $\times$$1^{''}$=1, 
\quad $1^{''}$$\times$$1^{''}$=$1^{'}$\\
3$\times$ 1=3, \quad 3$\times$$1^{'}$=3, \quad 3$\times$
  $1^{''}$=3.\\
\end{flushleft}
$A_4$ can also be generated by the two elements $S$ and $T$ 
satisfying the relation 
$S^2$ = ${(ST)}^3$ = $T^3$ =$1$ where S =(4321) and T = (2314). 
Under such representation the above mentioned decomposed states can be 
viewed as 
\begin{flushleft}
1 \quad S=1 \quad T=1\\
1' \quad S=1 \quad $T=e^{i4\pi/3}$ = $\omega^2$\\
1''\quad S =1 \quad $T=e^{i2\pi/3}$ = $\omega$\\
\end{flushleft}
$A_4$ has two subgroups $G_S$ : reflection subgroup generated by S isomorphic 
to $Z_2$ , $G_T$ : rotational subgroup generated by T and isomorphic to 
$Z_3$. If the flavor symmetry associated to $A_4$ is broken by the 
VEV of a triplet $\phi = (\phi_1,\phi_2,\phi_3)$ there are two interesting 
breaking patterns generated. We will see in the next section $G_S$ and $G_T$ 
play a very crucial role to generate appropriate neutrino and charged 
lepton mass matrices which after diagonalization lead to HPS mixing matrix 
exactly. 
\section{Altarelli-Feruglio Model}
The field content of the model and their representation content is shown 
in Table 1.
\begin{table}
\begin{center}
\begin{tabular}{|c|c|c|c|}
\hline
{\rm Lepton}& $SU(2)_L$ & $A_4$&\\
\hline
$(\nu_i, l_i)$&2&3&\\
$l_i^c$&1&1&\\
\hline
Scalar&&&{\rm VEV}\\
\hline
$h_u$&2&1&$<h_u^0>$= $v_u$\\
$h_d$&2&1&$<h_d^0>$=$v_d$\\
$\xi$&1&1&$<\xi^0>$ = u\\
$\phi_S$&1&3&$<\phi_S^0>$ = $(v_S,v_S,v_S)$\\
$\phi_T$&1&3&$<\phi_T>$ = $(v_T,0,0)$\\
\hline
\end{tabular}
\caption{List of fermion and scalar fields used in AF  model.
}
\end{center}
\end{table}
The most general $SU(2)_L\times U(1)_Y\times A_4$ symmetry invariant 
Yukawa interaction is given by \cite{Altarelli1}
\begin{eqnarray}
{{L}_l=y_ee^c(\phi_Tl)h_d/\Lambda+y_\mu\mu^c(\phi_Tl)'h_d/\Lambda
+y_\tau\tau^c(\phi_Tl)''h_d/\Lambda}\nonumber\\
{+x_a\xi(lh_ulh_u)/\Lambda^2+x_b(\phi_Slh_ulh_u)/\Lambda^2+h.c}
\end{eqnarray}
where $x_a$,$x_b$,$y_e$,$y_\mu$,$y_\tau$ are Yukawa couplings and 
$\Lambda$ is the scale of new physics. After spontaneous breaking of 
$A_4$ due to $<\phi_S>\neq 0$ and $<\phi_T>\neq 0$ and $SU(2)_L\times U(1)_Y$
due to $<h_u>\neq 0$ and $<h_d>\neq 0$, we obtain the charged lepton 
and neutrino mass matrices as 
\begin{equation}
m_l = \pmatrix{m_e&0&0\cr
               0&m_\mu&0\cr
               0&0&m_\tau},
m_\nu = \pmatrix{a + 2d/3&-d/3&-d/3\cr
                    -d/3&2d/3&a-d/3\cr
                    -d/3&a-d/3&2d/3},
\end{equation}
where
 ${m_e=y_ev_d\frac{v_T}{\Lambda}}$,
  ${m_\mu=y_\mu v_d\frac{v_T}{\Lambda}}$,
    ${m_e=y_\tau v_d\frac{v_T}{\Lambda}}$,
      ${a=2x_av_u^2\frac{u}{\Lambda^2}}$ and
      ${d=2x_bv_u^2\frac{v_S}{\Lambda^2}}$    
Upon diagonalization of $m_\nu$, the three non degenerate eigenvalues 
obtained as 
\begin{equation}
m_1 = a + d , \quad\quad m_2 = a, \quad\quad m_3 = d - a 
\end{equation}
and the mixing matrix comes out same as the tri-bimaximal pattern. It
is to be noted that choice of VEV is not very natural in the present
model that causes vacuum alignment problem, however, that has been
overcome through either by supersymmetrization 
\cite{Altarelli1,Altarelli2} of
the model or through the introduction of another extra space dimension
\cite{Altarelli3}.
\section{Beyond AF Model}
Although AF model model successfully generates tri-bimaximal 
mixing, however, due to the following reasons it is 
our endeavor to go beyond AF model: 
\begin{itemize}
\item  HPS mixing gives 
$\theta_{12} = 35.26^o$ which is at the 1$\sigma$ range of 
the solar angle whereas the best fit value is 33.9$^o$,
\item $\theta_{13}$ may be non-zero according to CHOOZ experimental
bound,
\item Non zero value of $\theta_{13}$ associated with Dirac phase 
could lead to CP violation in the leptonic sector which will be searched 
in the future short and long baseline experiments.
\end{itemize}

Motivated with the above, AF model has been modified through the 
inclusion of a SU(2)$_{\rm L}$ singlet charged scalar 
$\chi = (\chi_1,\,\, \chi_2,\,\, \chi_3)$ and due to the soft 
breaking of $A_4$ symmetry in the scalar sector. 
A trilinear $A_4$ symmetry breaking term is incorporated and due to 
radiative Zee mechanism the following mass matrix of the neutrino 
is obtained \cite{Adhikary:2006wi}
\begin{equation}
m_\nu = \left ( \begin{array}{ccc} a + \frac {2d} {3} & -\frac {d} {3} 
& -\frac {d} {3} - \epsilon \\
&& \\
-\frac {d} {3} & \frac {2d} {3} & a -\frac {d} {3} + \epsilon \\
&& \\
-\frac {d} {3} - \epsilon & a -\frac {d} {3} + \epsilon & \frac {2d} {3} 
\end{array} \right )
\end{equation}
where $\epsilon = f m_\tau^2 \frac {c_{12} v_u} {v_d} F(m_\chi^2, m_{h_d}^2)$
with 
\begin{equation}
F(M_1^2, M_2^2) = \frac {1} {16\pi^2 (M_1^2 - M_2^2)} \ln 
\left ( \frac {M_1^2} {M_2^2} \right ).
\end{equation}
Here we consider two cases:
\noindent {\bf i)} Here we assume $a$, $d$, $\epsilon$ are all real and 
$|\epsilon | << |a|, \,\, |d|$,

\noindent we obtain three eigenvalues as 
\begin{equation}
m_1 = a + d + \epsilon, \,\,\,\, m_2=a\, ,\,\,\,\, m_3 = d - a - \epsilon
\end{equation}
and the three mixing angles are 
\begin{equation}
\sin \theta_{12} = \frac {1} {\sqrt{3}} + \delta_1, \,\,\,
\sin \theta_{23} = -\frac {1} {\sqrt{2}} - \delta_2, \,\,\,
\sin \theta_{13} = \delta_3
\end{equation}
where
$$
\delta_1 = \frac {\epsilon} {d\sqrt{3}}, \,\,\,
\delta_2 = \frac{1}{3} \left [ \frac {\epsilon\sqrt{2}}{4a} - 
\frac {\epsilon}{\sqrt{2}(2a - d)} \right ], \,\,\,
\delta_3 = \frac{1} {3} \left [ \frac {\epsilon\sqrt{2}}{a} +
\frac {\epsilon}{\sqrt{2}(2a - d)} \right ].
$$
Assuming a constraint relation $d = -2a$, from Eq. (12) 
we obtain,
\begin{eqnarray}
m_1 &=& - \frac {1} {2\sqrt{2}} \sqrt{\Delta m^2_{\rm atm}} + 
\sqrt{2}\,\, \sqrt{\Delta m^2_{\rm atm}} R \nonumber \\
m_2 &=& \frac {1} {2\sqrt{2}} \sqrt{\Delta m^2_{\rm atm}} \nonumber \\
m_3 &=& - \frac {3} {2\sqrt{2}} \sqrt{\Delta m^2_{\rm atm}} -
\sqrt{2}\,\, \sqrt{\Delta m^2_{\rm atm}} R \nonumber
\end{eqnarray}
and 
\begin{eqnarray}
\sin\theta_{13} &=& \frac {5} {3\sqrt{2}}\,R  \nonumber \\
\sin\theta_{12} &=& \frac {1} {\sqrt 3} - \frac {2\sqrt{6}} {5} 
\sin\theta_{13} \nonumber \\
\tan^2\theta_{23} &=& 1 + \frac {4\sqrt{2}} {5} \sin\theta_{13}
\nonumber
\end{eqnarray}
where $R = \Delta m^2_\odot / \Delta m^2_{\rm atm}$.
Using $\Delta m^2_\odot = (7.2 - 8.9) \times 10^{-5}$ eV$^2$
$\Delta m^2_{\rm atm} = (1.7 - 3.3) \times 10^{-3}$ eV$^2$ and
$R = (2.2 - 5.2) \times 10^{-2}$ we have 
$$
m_1 = 0.015 {\rm eV}, \,\,\, m_2 = 0.017 {\rm eV}, \,\,\,m_1 = - 0.055 {\rm eV}
$$
$$
\theta_{12} =31.13^o - 33.5^o,\,\,\,\theta_{12} =43.5^o - 46^o,\,\,\,
\theta_{12} =3.5^o - 1.5^o
$$
Thus the mass eigenvalues are normally ordered. While there is hierarchy
between $m_{1,2}$ and $m_3$, the masses $m_1$ and $m_2$ are nearly 
quasi-degenerate. The above mass values also give
$$
|m_{ee}| \simeq 0.01 {\rm eV}
$$
which is far beyond the detection limits of planned $\beta\beta_{0\nu}$
experiments in near future. Furthermore, the smaller value of $\theta_{13}$
is correlated with the larger value of $\theta_{12}$ and $\theta_{23}$. 
For example, with $d= -2.25a$, we obtain $\theta_{13} = 6^o$, but 
$\theta_{12} = 29.2^o$ which is below the range allowed at 99.9$\%$ C.L. 
and hence ruled out.

\noindent {\bf ii)}  One Complex and two real parameters:

In this section we consider the above model with one complex and two 
real parameters which in addition to the three mixing angles and three 
mass eigenvalues give rise to CP violation in the leptonic sector 
which has been parametrized through the $J_{CP}$ parameter. We express 
 mixing angles $\theta_{13}$ and $\theta_{23}$ in terms of a single 
model parameter and constrained the allowed parameter space. Utilizing 
all those constraints, we explore the extent of $J_{CP}$. We also 
describe the mass pattern obtained in this case \cite{Adhikary:2006jx}. 

We consider the parameter `$d$' is complex and the neutrino mass matrix comes 
out as 
\begin{equation}
m_\nu = \left ( \begin{array}{ccc} 
a + \frac {2d e^{i\phi}} {3} & -\frac {d\,e^{i\phi}} {3} & 
-\frac {d\,e^{i\phi}} {3} - \epsilon \\
&& \\
-\frac {d\,e^{i\phi}} {3} & \frac {2d e^{i\phi}} {3} & 
a - \frac {d e^{i\phi}} {3} + \epsilon \\
&& \\
- \frac {d e^{i\phi}} {3} - \epsilon & a - \frac {d e^{i\phi}} {3} + \epsilon &
\frac {2d e^{i\phi}} {3} 
\end{array} \right )
\end{equation}
Diagonalizing the neutrino mass matrix through 
\begin{equation}
U^\dagger m_\nu U^* = {\rm diag} (d\,e^{i\phi} + a + \epsilon,\,\, a, \,\,
d\,e^{i\phi} - a - \epsilon)
\end{equation}
up to the first order of $\epsilon$ and the matrix $U$ becomes 
\begin{equation}
U = U_{tb} + {\cal O} (\epsilon) ... 
\end{equation}
Explicitly we get,
\begin{eqnarray}
\sin\theta_{12} = |U_{12}| & = & \frac {1} {\sqrt{3}} +  
\frac {\epsilon' (2 + \kappa \cos^2 \phi)} 
{\sqrt{3} \kappa \cos^2\phi (\kappa + 2)} \nonumber \\
\sin\theta_{13} = |U_{13}| & = & \left | \frac {\epsilon'} 
{3\sqrt{2} (\kappa - 2)} \right | 
\left [ \frac {1 + \cos^2\phi (\kappa^2 + 8 - \epsilon \kappa)}
{\cos^2\phi} \right ]^{1/2} \\
\tan^2\theta_{23} &=& 1 + \frac {2\epsilon' \kappa} {3(\kappa - 2)} \nonumber
\end{eqnarray}
where $\epsilon' = \epsilon/a, $\,\,\,$d= -\kappa a\cos\phi$.
It is possible to express the parameter $a^2$ as 
\begin{equation}
a^2 = \frac {\Delta m^2_{23} (2 + \kappa \cos^2\phi)}
{\kappa \cos^2\phi[(\kappa -2)(2+\kappa \cos^2\phi) + 
2\sqrt{3} (\kappa +2)
(1 - \kappa \cos^2\phi) (\sin\theta_{12} - \frac {1} {\sqrt{3}} ]} 
\end{equation}

From the above equation we get $\Delta m^2_{32} > 0$ since $a^2 > 0$. 
Also if we express $\cos^2\phi$ in the similar way, we get for the best fit 
value of observables $\kappa < -2$ for $\cos^2\phi < 1$. The model gives for 
$\theta_{12} = 34.0^o$, $\theta_{23} > 45.8^o$ and $\theta_{13} = 3.0^o$.
If we allow $1\sigma$ deviation of $\theta_{23}$ $(\theta_{23} = 48^o)$ 
the maximum possible allowed value obtained from the experiment
(CHOOZ : $0^o < \theta_{13} < 12^o$ at 3$\sigma$).
\par
Keeping all those constraints in view we explore the parameter 
space of CP violation parameter $J_{CP}$. The parameter $J_{CP}$
is defined as 
\begin{eqnarray}
J_{CP} =
  \frac{1}{8}\sin2\theta_{12}\sin2\theta_{23}\sin2\theta_{13}
  \cos\theta_{13}\sin\delta_D
   = \frac{Im[h_{12}h_{23}h_{31}]}
   {\Delta m^2_{21}\Delta m^2_{31}\Delta m^2_{32}}
\end{eqnarray}
where $h= m_\nu m_\nu^\dagger$. From the analytical expression of 
$J_{CP}$ it has been found that $1^o$ deviation of 
$\theta_{23}$($46^o$ at $\kappa= -2.03$) from its best fit value 
predicts $J_{CP}$ = $7.1\times 10^{-5}$. If we allow 
$1\sigma$ deviation of $\theta_{23}$ 
($\theta_{23} = 48^o$ at $\kappa = -2.41$) the value of 
$J_{CP}$ = $2.6\times {10}^{-3}$ is larger and which can be 
probed through the up coming base line experiments. 
\par 
It is also possible to constrain $\delta_D$ and for the best-fit 
values of $\theta_{12}$, $\Delta m^2_{21}$, $\Delta m^2_{32}$ 
we obtain $\delta_D$ = $3.6^o$ for $\theta_{23} = 48^o$ at 
$\kappa = -2.41$. Moreover, the sum of the three neutrino masses vary 
as $0.07<m_1+m_2+m_3<0.076$ due to the variation of parameter 
$\kappa$, $-2.41<\kappa<-2$ ($48^o>\theta_{23}>45.8^o$) which is 
also consistent with the cosmological bounds on the sum of the neutrino mass.  
\section{Open Problems}
There are still few open problems which needs further attention \\
1. Embedding within a GUT model, so that the whole scenario 
of quarks and leptons could be explained. One of the major 
ongoing problem is to achieve quark-lepton complementarity 
relation which gives $\theta_c + \theta_{12} = 45.1^o \pm 2.4^o(1\sigma)$
\cite{cmpl}. An attempt has been made within the context of supersymmetric
$SU(4)_c\times SU(2)_L\times SU(2)_R$ model \cite{pati}, however, a successful 
scenario with $SU(2)_L\times U(1)_Y\times A_4$ model is still 
yet unanswered.\\
2. See-saw realization of the AF model has been 
done \cite{Altarelli1,seesaw}, however, 
connection between CP violating phases with leptogenesis 
has not yet been studied. 
\section{Acknowledgement}
Author thanks the organizers of the XVII DAE-BRNS High Energy Physics 
Symposium held at Indian Institute of Technology, Kharagpur, India 
 during December 
2006, for the invitation 
to give this talk. Author also acknowledges Biswajit Adhikary 
for many helpful discussions.

\end{document}